\documentclass[10pt, conference]{IEEEtran}

\usepackage{mystyle}

\def\doctitle{APIzation: Generating Reusable APIs from \\ StackOverflow Code Snippets}
\StrSubstitute{\doctitle}{ \\ }{ }[\cleandoctitle]

\def\docauthors{Valerio Terragni, Pasquale Salza}

\def\dockeywords{%
APIs, software reuse, code snippets, StackOverflow, GitHub, program analysis, program synthesis
}

\hypersetup{
	pdftitle={\cleandoctitle},
	pdfauthor={\docauthors},
	pdfkeywords={\dockeywords}
}

\DeclareAcronym{api}{
	short = API,
	long = {Application Program Interface}
}

\DeclareAcronym{cnn}{
	short = CNN,
	long = {Convolutional Neural Network}
}

\DeclareAcronym{cs}{
	short = CS,
	long = {Code Snippet}
}

\DeclareAcronym{dl}{
	short = DL,
	long = {Deep Learning}
}

\DeclareAcronym{id}{
	short = ID,
	long = {Identifier}
}

\DeclareAcronym{jd}{
	short = JD,
	long = {Jaccard Distance}
}

\DeclareAcronym{ld}{
	short = LD,
	long = {Levenshtein Distance}
}

\DeclareAcronym{lstm}{
	short = LSTM,
	long = {Long Short Term Memory}
}

\DeclareAcronym{pos}{
	short = POS,
	long = {Part-of-Speech}
}

\DeclareAcronym{rnn}{
	short = RNN,
	long = {Recurrent Neural Network}
}

\DeclareAcronym{so}{
	short = SO,
	long = {StackOverflow}
}

\DeclareAcronym{wmd}{
	short = WMD,
	long = {Word Mover's Distance}
}

\DeclareAcronym{gh}{
	short = GH,
	long = {GitHub}
}

\DeclareAcronym{loc}{
	short = LOC,
	long = {Lines of Code}
}

\DeclareAcronym{ast}{
	short = AST,
	long = {Abstract Syntax Tree}
}

\DeclareAcronym{jdk}{
	short = JDK,
	long = {Java Development Kit}
}

\begin{document}

\title{\doctitle}

\author{

\IEEEauthorblockN{Valerio Terragni}
\IEEEauthorblockA{%
University of Auckland\\
Auckland, New Zealand\\
v.terragni@auckland.ac.nz%
}

\and

\IEEEauthorblockN{Pasquale Salza}
\IEEEauthorblockA{%
University of Zurich\\
Zurich, Switzerland\\
salza@ifi.uzh.ch%
}

}

\maketitle

\begin{abstract}
Developer forums like StackOverflow have become essential resources to modern software development practices.
However, many code snippets lack a well-defined method declaration, and thus they are often incomplete for immediate reuse.
Developers must adapt the retrieved code snippets by parameterizing the variables involved and identifying the return value. %
This activity, which we call \apization of a code snippet, can be tedious and time-consuming.
In this paper, we present \tool to perform \apizations of \java code snippets automatically.
\tool is grounded by four common patterns that we extracted by studying real \apizations in GitHub.
\tool presents a static analysis algorithm that 
automatically extracts the method parameters and return statements.
We evaluated \tool with a ground-truth of \num{200} \apizations collected from \num{20} developers.
For \num{113} (\SI{56.50}{\percent}) and \num{115} (\SI{57.50}{\percent})
\apizations, \tool and the developers extracted identical parameters and return statements, respectively.
For \num{163} (\SI{81.50}{\percent}) \apizations, either the parameters or the return statements were identical.

\end{abstract}


\section{Introduction}
\label{sec:introduction}

\watermark{%
This is the authors' version of the paper that has been accepted for publication in the\\%
36th IEEE/ACM International Conference on Automated Software Engineering (ASE 2021)%
}
Developers' \qa websites, such as \acf{so}, have gained a lot of popularity.
These websites contain millions of crowd-curated code snippets that represent solutions to various programming tasks.
These code snippets are extremely useful to both developers and researchers.
Developers often search for them to draw inspiration or simply reuse them~\cite{brandt_examplecentric_2010,storey_evolution_2014,mao_survey_2017}. 
Researchers often rely on \ac{so} to accomplish various software engineering goals~\cite{meldrum_crowdsourced_2017}.

\begin{figure*}[!tb]
	\centering
	\includegraphics[width=0.99\linewidth]{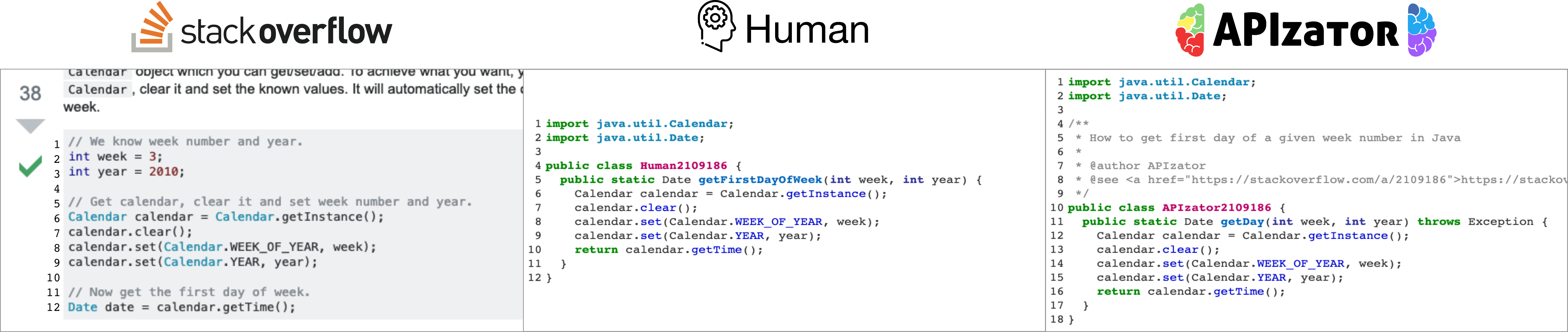}
	\caption{\apization of a \ac{so} code snippet. \tool and the human %
	produced identical \apis (except for the method name and \textsc{JavaDoc}).%
	}
	\label{fig:example_first_day}
\end{figure*}

When reusing \ac{so} code snippets, developers and researchers face a major obstacle: most \ac{so} code snippets do not compile~\cite{terragni_csnippex_2016,subramanian_making_2013,yang_query_2016}.
It mainly occurs because they are written for illustrative purposes, to convey solutions at a high level, without implementation details~\cite{nasehi_what_2012}.
Terragni \etal have shown that $\approx$\SI{92}{\percent} of \num{491906} analyzed \ac{so} code snippets are uncompilable~\cite{terragni_csnippex_2016}.
A common missing implementation detail is the type declaration~\cite{terragni_csnippex_2016,subramanian_making_2013}.
For instance, the \java \ac{so} code snippet in \cref{fig:example_first_day} (left side) misses the declaration of type \code{Calendar} and \code{Date}.
Researchers have tackled this issue by proposing techniques to identify the import declarations required to compile \ac{so} code snippets~\cite{terragni_csnippex_2016,phan_statistical_2018}.

Another common missing detail in \ac{so} code snippets is a well-formed method declaration that defines the parameters (input) and return statements (output)~\cite{treude_understanding_2017,terragni_csnippex_2016}.
Terragni \etal have shown that~$\approx$\SI{56}{\percent} of \java \ac{so} code snippets constitute \emph{dangling statements}, which are not embedded in any class or method declarations~\cite{terragni_csnippex_2016}. 
The \ac{so} code snippet in \cref{fig:example_first_day} (left side) is an example of dangling statements. 
One could automatically wrap the code snippet inside a generic method declaration~\cite{terragni_csnippex_2016,subramanian_making_2013} (\eg, the \code{main} function).
It would resolve compilation errors but would not recover the proper method declaration that exposes the intended input and output of the code snippet.
The absence of a proper interface prevents the direct reuse of \ac{so} code snippets.
Thus, some manual effort is required to identify the inputs and outputs of the code snippets.

We use the term \emph{\enquote{\apization{}}} to indicate the activity of creating an \acf{api} for those \ac{so} code snippets without a well-formed method declaration.
\Cref{fig:example_first_day} (center) shows a  manual \apization of a \ac{so} code snippet.

In this paper, we study the automatic \apization of \java \ac{so} code snippets, which would bring important benefits.
Developers would reduce the effort of integrating \ac{so} code snippets into their codebases, which is known to be a tedious and time-consuming activity~\cite{holmes_systematizing_2012}.
Given an automatically generated \api of a \ac{so} code snippet, developers can either copy and paste the \api in the codebase or incorporate the method body of the \api inside an existing method.
The presence of an \api facilitates the latter option.
Indeed, an \api explicitly shows the input and output of the code snippet, which helps to both understand and incorporate the \ac{so} code. %
Moreover, the automatic \apization \ac{so} code snippets can lead to a large catalog of code samples with well-defined interfaces, providing value for both developers and researchers. 

Towards these goals, we conducted an investigatory study to understand how developers perform \apizations from \ac{so} code snippets to \java methods found in \acf{gh}.
The insights gained from this study led to four common \apization patterns to extract method parameters and return statements.
Grounded by these patterns, we propose a technique called \tool for the automated \apization of \ac{so} code snippets.
To the best of our knowledge, \tool is the first technique of its kind.
\tool statically analyzes a given code snippet to find matches for the four patterns.
If it finds matches, \tool extracts the parameters and return statements and outputs a compilable \api.
For completeness, \tool uses a \ac{pos} Tagger to generate a method name from the \ac{so} question title, and creates a \textsc{JavaDoc} containing the title and link of the corresponding \ac{so} page.

We evaluated \tool with a ground truth of \num{200} \apizations performed by \num{20} human participants, obtaining \num{200} pairs of human- and tool-produced \apis.
We compared each pair to assess the effectiveness of \tool.
For \num{113} (\SI{56.50}{\percent}) and \num{115} (\SI{57.50}{\percent}) \api pairs the parameter list and return statements are identical, respectively.
For \num{163} (\SI{81.50}{\percent}) \apis generated by \tool either the return statements or the method parameters are identical to those produced by the developers.
For instance, \cref{fig:example_first_day} (right side) shows the \api produced by \tool, which is identical to the one created by the developer (excluding the method name and \textsc{JavaDoc}).

To demonstrate one of the possible usage scenarios of \tool, we release a search engine at the address \url{https://apization.netlify.app/search/} and as part of our replication package~\cite{replicationpackage}.
The users can search for \ac{so} code snippets with a natural language query as they would do with a standard search engine.
The search results show the \ac{so} page as well as its \api automatically generated by \tool.

To summarize, the main contributions of this paper are:
\begin{itemize}
	\item studying the problem of automatically transforming \ac{so} code snippets into \apis;
	\item analyzing real \apizations across \ac{so} and \ac{gh} projects, extracting four common \apization patterns;
	\item proposing a technique called \tool to transform \ac{so} code snippets into well-formed \java method declarations; %
	\item evaluating \tool against a ground truth of \num{200} \apizations performed by \num{20} \java developers;
	\item releasing at the address \url{https://apization.netlify.app} all the experimental data;
	\item releasing \num{109930} \apis automatically extracted from \ac{so} code snippets, which could power \ac{so}-centric research.
\end{itemize}

\section{Preliminaries and Problem Definition}
\label{sec:background}

In this paper, we target \java code snippets found in \acf{so}, the most popular \qa website for developers~\cite{vasilescu_how_2014}.
The process of \apization takes in input a \ac{so} code snippet and generates a \java method declaration. %
We now describe in detail the input and output of such a process.

\paragraph{Input: A \java code snippet from \ac{so}}
A \ac{so} page is composed of a question post and a series of answer posts.
Each question post contains a title, a series of tags, and a description.
A post can contain one or more code snippets.
A \textbf{\ac{cs}} is an ordered sequence of source code lines.

\paragraph{Output: A compilable and well-formed \java method declaration that defines the code snippet in input}
A \emph{method declaration}, which we call \acf{api}, consists of the following six attributes:
\begin{inparaenum} [(i)]
	\item \emph{modifiers}, which set the access level (\eg, \code{public}), or achieve specific functionalities (\eg, \code{static});
	\item \emph{return type}, which indicates the type of value that the method returns (\code{void}  if none); %
	\item \emph{method name}, which  describes the behavior of the method;
	\item \emph{parameter list}, which specifies the types and identifiers of the method arguments;
	\item \emph{throws clause}, which declares any checked exception classes that the method body can throw;
	\item \emph{method body}, which contains the statements that implement the method.
\end{inparaenum}

The \emph{method body} of a \emph{well-formed} \ac{api} references each of the parameters and contains, if the return type is not \code{void}, one or more return statements.
To make an \api compilable, it has to be declared inside a class (\eg, \code{Human2109186} in \cref{fig:example_first_day}) that contains the required \emph{import declarations} (\emph{imports} in short) (\eg, \code{java.util.Calendar} and \code{java.util.Date} in \cref{fig:example_first_day}).
At each class is associated a \emph{classpath} to the library \textsc{JARs} that declare the types in \emph{imports} (\eg, \acs{jdk} in \cref{fig:example_first_day}).

Most \java code snippets from \ac{so} are composed of dangling statements not enclosed in any method declaration~\cite{terragni_csnippex_2016, ponzanelli_improving_2014, subramanian_making_2013} (see \cref{fig:example_first_day}).
The \textbf{process of \apization} aims at generating well-formed method declarations for such code snippets.
It achieves this by performing six actions:
\begin{enumerate}
	\item choose a method name, \eg, \code{getFirstDayOfWeek} in \cref{fig:example_first_day};
	\item recover missing declarations of variables or types from the code snippet, \eg, \code{Calendar} and \code{Date} in \cref{fig:example_first_day};
	\item identify which variables in the snippet are the intended input parameters, \eg, variables \code{week} and \code{year} in \cref{fig:example_first_day};
	\item remove the declarations of such variable from the code snippet, \eg, \code{int week = 3;} in \cref{fig:example_first_day};
	\item infer the output of the  snippet, if any, and add a return statement for it, \eg, \code{return calendar.getTime()} in \cref{fig:example_first_day};
	\item enclose the resulting statements in a method declaration with proper parameters and return type, \eg, \code{public static Date (int week, int year)} in \cref{fig:example_first_day}.
\end{enumerate}

\begin{custombox}{Problem definition}
	Given a \java code snippet, the process of \textbf{\apization} generates a compilable and well-formed method declaration for the given code snippet. %
\end{custombox}

\section{Understanding Real World \apizations}
\label{sec:study}

\begin{figure*}[!tb]
	\centering
	\includegraphics[width=0.8\linewidth]{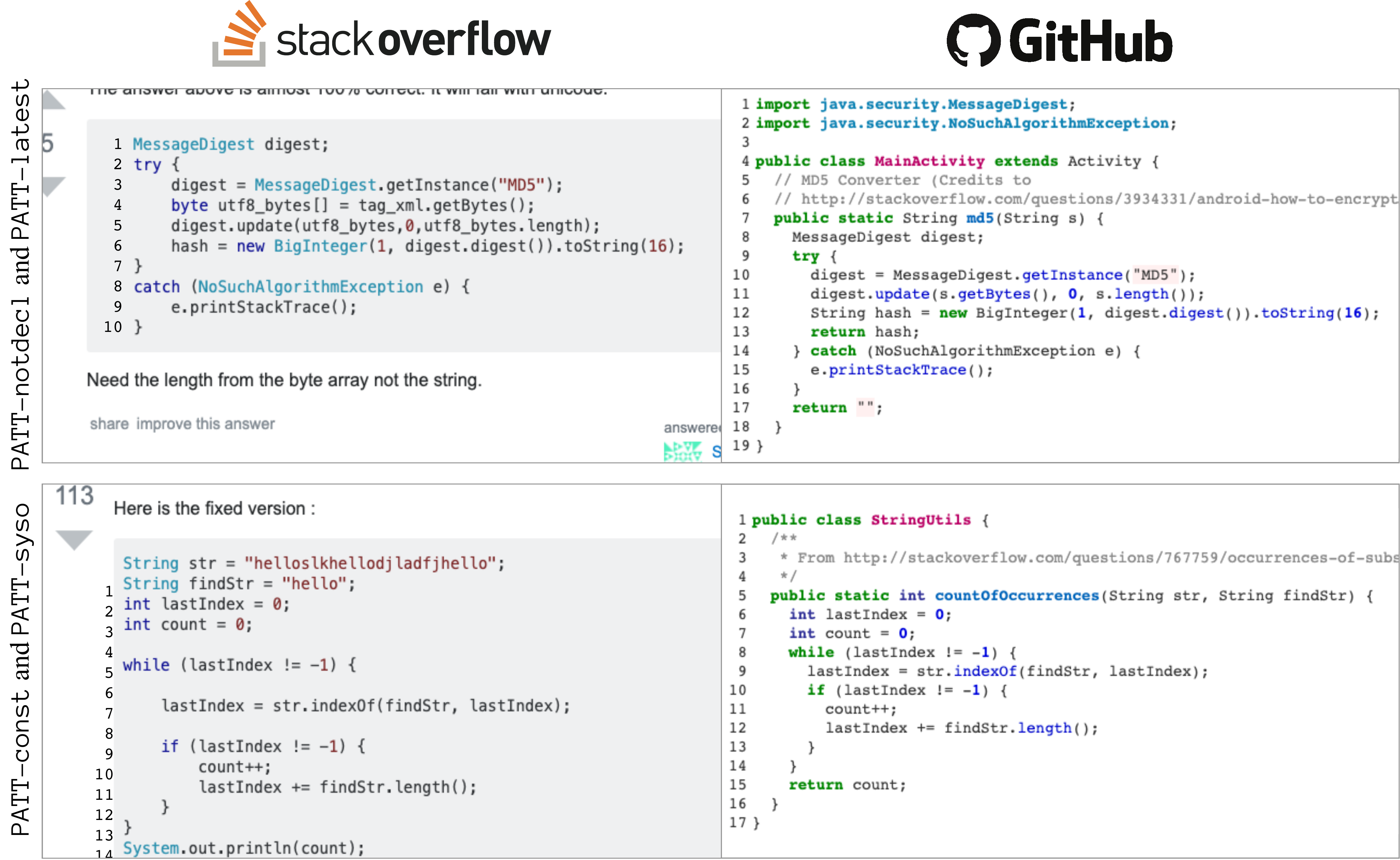}
	\caption{Examples of \apization patterns found in \apizations from \acl{so} to \acl{gh}.}
	\label{fig:study:examples_sogh}
\end{figure*}

This section presents an investigatory study to understand how developers perform \apizations.
The insights gained from this study led to four common \apization patterns that establish the foundations of our proposed technique.
To collect manual \apizations of \acf{so} code snippets, we relied on \acf{gh}. %
Our goal is to find code reuses across \ac{so} code snippets and \ac{gh} projects that represent \apizations.
\Cref{fig:study:examples_sogh} gives two examples of such manual \apizations.
We release the data of our investigatory study in our replication package, published at \url{https://apization.netlify.app/study/}.

\subsection{Data Collection}

Researchers have experimented two main approaches to identify code reuses across \ac{so} and \ac{gh}~\cite{gharehyazie_here_2017,vasilescu_stackoverflow_2013,badashian_involvement_2014,baltes_attribution_2017,zhang_are_2018,baltes_usage_2019,zhang_are_2018}:
\begin{inparaenum}[(i)]
	\item search for explicit \ac{so} web links in \ac{gh} code comments or \textsc{Javadoc};
	\item search for code clones between \ac{so} code snippets and \ac{gh} code.
\end{inparaenum}

Both of these approaches have pros and cons.
Relying only on explicit \ac{so} web links likely misses many code reuses.
In fact, \ac{gh} developers often do not give proper credit when reusing \ac{so} code snippets~\cite{baltes_attribution_2017,baltes_usage_2019}.
It can also lead to spurious code reuses as \ac{gh} developers may cite a \ac{so} post because it discusses a particular issue, which is unrelated to code reuse~\cite{baltes_usage_2019}. %
Relying only on code clones has the advantage to identify code reuses even without (rare) explicit \ac{so} links.
However, code clones cannot guarantee that the \ac{gh} developers performed the \apization from \ac{so}~\cite{baltes_usage_2019,gharehyazie_here_2017}.

Because of the complementarity of these two approaches, we decided to consider those code reuses that are identified by both approaches.
We will probably miss many code reuses, but we are more confident that the identified ones are genuine.
Thus, our goal is to identify pairs $\langle\textit{\cs}, \textit{\api}\rangle$ (where \textit{\cs} is a \ac{so} code snippet and \textit{\api} a \ac{gh} method) that satisfy all of these three criteria:
\begin{inparaenum}[(i)]
	\item the comments or \textsc{Javadoc} of \api have an explicit link to the \ac{so} page containing \ac{cs};
	\item \api and \ac{cs} are code clones;
	\item \api is an \apization of \ac{cs}.
\end{inparaenum}
We now describe in more detail how we identified such pairs.

\paragraph{Find candidate pairs} %
We queried the
latest
snapshot of \ac{gh} on \textsc{Google BigQuery}~\cite{google_bigquery}, which contains $\approx$\num{1}~million projects with the tag \java.
We identified %
\num{29035} unique \java files containing explicit links to \ac{so} pages.
From the retrieved Java files, we identified all the \ac{gh} methods (\api) containing the explicit \ac{so} link as a code comment or in the JavaDoc.
For each \ac{so} link, we extracted the corresponding \ac{so} code snippet(s) by querying the latest \ac{so} dump.

We then pruned all those pairs in which \textit{\cs} already contains a well-formed method declaration, or 
\textit{\cs} has less than six lines. %

\paragraph{Code clone detection}
For each candidate pairs $\langle\textit{\cs}, \textit{\api}\rangle$, we searched for \emph{TYPE~3} code clones~\cite{kamiya_ccfinder_2002}, \ie, syntactically similar code with inserted, deleted, or updated statements.
We chose TYPE~3 clones because both \emph{TYPE~2} and \emph{TYPE~4} are inadequate for our purposes.
\emph{TYPE~2} clones require syntactically equivalent code (the only allowed variations are in identifiers, types, whitespace, layout, and comments).
This is too restrictive because the \apizations often create \apis with fewer or more statements than the \ac{so} code snippets.
For example, the human \apization of \cref{fig:example_first_day} deletes the \ac{so} lines 2 and 3 and updates line 12.
TYPE~4 clones allow semantically equivalent but syntactically different code.
This is too permissive because we are only interested in explicit code reuses. %

To detect TYPE~3 clones, we automatically perform \emph{alpha-renaming} of the variables (\eg, \code{int a = 5} becomes \code{int int0 = 5}).
If there are multiple variables with the same type, we use a progressive id as a suffix.
For example, \code{int a = 5; int b = 10} becomes \code{int int0 = 5; int int1 = 10}.
We also removed comments, new lines, and formatting characters.
We treated a pair $\langle\textit{\cs}, \textit{\api}\rangle$ as a TYPE~3 code clone if at least \SI{70}{\percent} of \textit{\cs} source code lines are contained in \api (we opted for~\SI{70}{\percent} following Zhang \etal~\cite{zhang_are_2018}).
This resulted in \num{330} code clone pairs, referring to \num{199} unique \ac{so} answer posts.

Note that TYPE~3 code clone detection excludes by default TYPE~1 and TYPE~2 clones as they require a \SI{100}{\percent} similar code. This is impossible in our case since \apis always contain a method signature, while the considered code snippets do not.

\paragraph{Manual check}
We manually checked each of the \num{330} code clone pairs to prune those in which the \apis do not represent the \apization of \cs.
We pruned the pairs that were spurious code clones (the matched lines were mostly common lines of code such as \code{try\{} and \code{catch(}).
We pruned the pairs that were valid clones, but \textit{\cs} was incorporated inside the \ac{gh} method.
These pairs are not \apizations because the \ac{gh} method declaration does not strictly relate to the \ac{so} code snippet.

\subsection{Analysis of the Results}
\label{subsec:study:patterns}

\begin{revised}
We manually analyzed the retained \textbf{\num{135} pairs} to study the variables in the \ac{so} code snippet that became method parameters or return statements in the \ac{gh} method.
We followed a coding process inspired by \emph{grounded theory}~\cite{corbin_grounded_1990}, which derives new theories and concepts by analyzing the data.

We distributed between the two of us the \num{135} pairs of the \ac{so} snippet and matching \ac{gh} method.
For convenience, we used a diff tool to generate a visual representation of the code differences between the snippet and method.
Such a representation helped us to quickly identify the \apization activity performed by the developers.
During the \emph{open coding} stage, we analyzed each of the assigned pairs to give a distinct \emph{code} for each of the observed phenomena, \ie, \apizations.
In particular, the question that drove the open coding was: \enquote{What are the characteristics of the variables in the \ac{so} snippet that became parameters and return statements in the \ac{gh} method?}
Examples of produced codes are: \enquote{undeclared variable}, \enquote{the variable has a constant value}, and \enquote{the variable is used as an argument in a \code{System.out.println} invocation.}

Then, we refined the codes by grouping similar concepts and finding connections between them, \ie, \emph{axial coding}.
Then, we concluded the patterns' identification with \emph{selective coding}.

Each of the authors independently analyzed the pairs and eventually discussed the results to reach a consensus.
Finally, we identified four common patterns (\puno{}, \pdue{}, \ptre{}, and \pquattro{}) that characterize and define general \apization activities.
\end{revised}

\subsubsection{Method Parameters}

The \num{135} \ac{so} code snippets reference \num{509} variables with an average of \num{3.77} variables per code snippet.
Among these \num{509} variables, \num{45} became method parameters in the corresponding \ac{gh} method.
Among these \num{45} variables, \num{32} (\SI{71.11}{\percent}) match \puno{} and \num{9} (\SI{20.00}{\percent}) match \pdue{}.
For the remaining four variables, we were not able to generalize any pattern.

\paragraph{\puno{}}
\textit{A variable $v$ that is referenced in \textit{\cs} is extracted as a parameter if \textit{\cs} lacks the declaration of $v$.}

\Cref{fig:study:examples_sogh} (top) shows one of the analyzed pairs that exhibits such a pattern (the \ac{so} code snippet~\textit{(\cs)} is on the left and the \ac{gh} method~(\textit{\api}) on the right).
The line~4 of the \cs references an undeclared variable \code{tag\_xml}, and the \ac{gh} developer extracted \code{tag\_xml} as a method parameter at line~7 (renaming it to \code{s}).
A possible rationale for this pattern is that undeclared variables in \ac{so} code snippets are commonly intended as the (implicit) inputs of a programming task.

\paragraph{\pdue{}}
\textit{A variable $v$ declared in \textit{\cs} is extracted as a parameter if
\begin{inparaenum}[(i)]
	\item \textit{\cs} initializes $v$ with a hard-coded value; and
	\item \textit{\cs} does not have loops that modify the value of $v$.
\end{inparaenum}}

\Cref{fig:study:examples_sogh} (bottom) shows a pair that manifests such a pattern.
The \ac{so} code snippet declares four variables: \code{str}, \code{findStr}, \code{lastIndex}, and \code{count}.
It initializes them with hard-coded values that embed data directly into the source code.
These four variables match criterion (i), but only \code{str} and \code{findStr} match also criterion (ii).
In fact, only \code{str} and \code{findStr} became method parameters in the \ac{gh} method.
The variables \code{lastIndex} and \code{count} are excluded because the 
\ac{so} while loop can modify their values.
Extracting such variables would change the semantics of the while loop.
For example, if \code{count} is extracted as a parameter, a user can invoke the \api with a \code{count} value different from zero, making the \api return a meaningless value. %
A possible rationale for this pattern is that \ac{so} code snippets often exemplify programming tasks, and thus the hard-coded values represent a particular instance of the inputs. %

\subsubsection{Return Statements}

Among the \num{135} \ac{gh} methods, \num{63} (\SI{46.67}{\percent}) lack return statement(s) (the return type is \code{void}) and \num{72} (\SI{53.33}{\percent}) have return statement(s).
Among such \num{72} \ac{gh} methods, \num{31} (\SI{43.06}{\percent}) match \ptre{}, and \num{6} (\SI{8.33}{\percent}) match \pquattro{}.
For the remaining methods, we could not generalize any pattern or the \ac{so} code snippet already contained return statement(s).

\paragraph{\ptre{}}
\textit{The assignment of a variable in \textit{\cs} becomes the return statement if it is the last statement in \textit{\cs}.}

For example, the \ac{so} snippet in \cref{fig:study:examples_sogh} (top) ends with the assignment of the \code{hash} variable (we ignored exception handling as last statements because they are unrelated to the semantics of the code snippet), and the \ac{gh} method returns \code{hash} of type \code{String}.
Intuitively, the last statement of a \ac{so} snippet often characterizes its output.
Indeed, it is unlikely that developers end the snippet with a value irrelevant to the final intent of the programming task.

\paragraph{\pquattro{}}
\textit{If the last statement in \textit{\cs} is a \code{System.out.println} call, its argument becomes the return statement.}

An example of such a pattern is the \ac{so} snippet in \cref{fig:study:examples_sogh} (bottom).
The code snippet ends with \code{System.out.println(count)}, and the \ac{gh} method returns \code{count} of type \code{int}.
Because \ac{so} users write code snippets for illustration purposes, they often add a print of the output value to show the result when the snippet is being executed.

\subsubsection{Manual Application of the Patterns}
After identifying the four patterns, we applied them to the whole dataset to evaluate if they lead to spurious parameters and return statements.
Among the \num{464} \ac{so} variables that did not become parameters in the corresponding \ac{gh} methods,  \num{14} (\SI{3.02}{\percent}) and \num{8} (\SI{1.72}{\percent}) variables match \puno{} and \pdue{}, respectively.
Among the \num{93} GH methods in which we did not identify any pattern or lack return statements (i.e., return type is \code{void}), the patterns \ptre{} and \pquattro{} match \num{4} (\SI{4.30}{\percent}) and \num{1} (\SI{1.08}{\percent}) variables, respectively. 

This indicates that the four patterns lead to a few spurious parameters and return statements.
Thus, finding matches of these patterns in \ac{so} code snippets is a viable solution for automating the \apization process.

\section{\tool}
\label{sec:approach}

This paper presents \tool to automatically transform \java \ac{so} code snippets into reusable and compilable \apis.

\Cref{alg:approach} describes the process of \tool in detail.

\begin{algorithm}[!tb]
	\caption{\tool}
	\label{alg:approach}
	\footnotesize
	
\SetKwFunction{GetOrDefault}{GetOrDefault}

\SetKwFunction{Compile}{Compile}

\SetKwFunction{CreateInitialMethodDeclaration}{CreateInitialMethodDeclaration}
\SetKwFunction{RecoverVarType}{RecoverVarType}
\SetKwFunction{CSnippEx}{CSnippEx}
\SetKwFunction{CreateMethodName}{CreateMethodName}
\SetKwFunction{GetLoopChangingVars}{GetLoopChangingVars}
\SetKwFunction{GetVariable}{GetVariable}
\SetKwFunction{GetTypeOfExp}{GetTypeOfExp}
\SetKwFunction{IsHardCoded}{IsHardCoded}
\SetKwFunction{Type}{Type}

\Input{%
	\algovar{CS} $= \langle$\algovar{$s_1$}, $\dots$, \algovar{$s_n$}$\rangle$, a \algovar{SO} code snippet\newline
	\algovar{SO-page}, the SO page of \algovar{CS}\newline
    \algovar{JARs}, a set of external libraries
}
\Output{%
	\algovar{API}, a method declaration for \algovar{CS}\newline
	\algovar{imports}, the import declarations for \algovar{API}\newline
	\algovar{classpath}, the classpath for \algovar{API}
}
\BlankLine

$\langle$\algovar{imports}, \algovar{classpath}$\rangle$ $\leftarrow$ \GetOrDefault{\algovar{CS}, \algovar{JARs}}\label{alg:approach:getordefault}\;

\If{\upshape \algovar{CS} is a well-formed method declaration (\algovar{CS} $\equiv$ \algovar{API})}{\label{alg:approach:wellformed:condition}
	\Return{$\langle$\algovar{API}, \algovar{imports}, \algovar{classpath}$\rangle$}\label{alg:approach:wellformed:return}\;
}

\algovar{API} $\leftarrow$ \CreateInitialMethodDeclaration{\algovar{imports}, \algovar{CS}}\label{alg:approach:init}\;

\algovar{API.method-name} $\leftarrow$ \CreateMethodName{\algovar{SO-page}}\label{alg:approach:namegen}\;

\While{\Compile{\algovar{API}, \algovar{imports}, \algovar{classpath}} $\rightarrow$ \algovar{errors} $\neq \varnothing$}{\label{alg:approach:compile}
	\If{\upshape \algovar{errors} $\subseteq$ \algoval{missing-type-decl}}{\label{alg:approach:compile:misstypedecl}
		$\langle$\algovar{imports}, \algovar{classpath}$\rangle \leftarrow$ \CSnippEx{\algovar{errors}, \algovar{JARs}, \algovar{imports}, \algovar{classpath}}\label{alg:approach:compile:misstypedecl:end}\;
	}
	
	\tcc{\hfill\textbf{\puno{}}}
	\ElseIf{\upshape \algovar{errors} $\subseteq$ \algoval{missing-variable-decl}}{\label{alg:approach:compile:missvardecl}
		\For{\algovar{$v$} $\in$ (\algovar{errors} $\cap$ \algoval{missing-variable-decl})}{
			$\langle$\algovar{$\tau$}, \algovar{imports}, \algovar{classpath}$\rangle \leftarrow$ \RecoverVarType{\algovar{$v$}, \algovar{API}, \algovar{JARs}, \algovar{imports}, \algovar{classpath}}\label{alg:approach:compile:missvardecl:recovervartype}\;
			\algovar{$\mathcal{T}[v]$} $\leftarrow$ \algovar{$\tau$}\;
			add $\langle$\algovar{$\tau$}, \algovar{$v$}$\rangle$ to \algovar{API.parameter-list}\label{alg:approach:compile:missvardecl:addparam}\;
		}
	}
	\lElse{
		\Return{$\varnothing$}\label{alg:approach:compile:else}
	}
}

\tcc{\hfill\textbf{\pdue{}}}
\algovar{LP-VARS} $\leftarrow$ \GetLoopChangingVars{\algovar{API.method-body}}\label{alg:approach:hardcodeinit:getloopchangingvars}\;
\For{\algovar{$s_i$} $\in$ \algovar{API.method-body}}{\label{alg:approach:hardcodeinit:forbody}
	\Case(\hfill{\commentfont// Variable decl. and init.}){\algovar{$s_i$} : \algovar{$\tau$}~\algovar{$v$} $=$ \algovar{$\epsilon$}}{
		\algovar{$\mathcal{T}[v]$} $\leftarrow$ \algovar{$\tau$}\;
		add \algovar{$v$} to \algovar{ALREADY-INIT-VARS}\;
		\If{\upshape \IsHardCoded{\algovar{$\tau$}, \algovar{$\epsilon$}} $\land$ \algovar{$v$} $\nsubseteq$ \algovar{LP-VARS}}{\label{alg:approach:hardcodeinit:ishardcoded}
			add $\langle$\algovar{$\tau$}, \algovar{$v$}$\rangle$ to \algovar{API.parameter-list}\;
			remove \algovar{$s_i$} from \algovar{API.method-body}\;
		}
	}

	\Case(\hfill{\commentfont// Variable declaration}){\algovar{$s_i$} : \algovar{$\tau$}~\algovar{$v$}}{ 
		$\langle$\algovar{$\mathcal{T}[v]$}, \algovar{$\mathcal{S}[v]$}$\rangle$ $\leftarrow \langle$\algovar{$\tau$}, \algovar{$s_i$}$\rangle$\;
	}

	\Case(\hfill{\commentfont// Variable assignment}){\algovar{$s_i$} : \algovar{$v$} $=$ \algovar{$\epsilon$}}{ 
		\If{\upshape \algovar{$v$} $\not\in$ \algovar{ALREADY-INIT-VARS}}{\label{alg:approach:hardcodeinit:alreadyinitvars}
			add \algovar{$v$} to \algovar{ALREADY-INIT-VARS}\; 
			\If{\upshape \IsHardCoded{\algovar{$\tau$}, \algovar{$\epsilon$}} $\land$ \algovar{$v$} $\not\in$ \algovar{LP-VARS}}{
				add $\langle$\algovar{$\mathcal{T}[\tau]$}, \algovar{$v$}$\rangle$ to \algovar{API.parameter-list}\;
				remove \algovar{$s_i$} from \algovar{API.method-body}\;
				remove \algovar{$\mathcal{S}[v]$} from \algovar{API.method-body}\label{alg:approach:hardcodeinit:removebody}\;
			}
		}
	}
}

\tcc{\hfill\textbf{\ptre{}}}
\Case(\hfill{\commentfont// Variable decl. and init.}){\algovar{$s_n$} : \algovar{$\tau$}~\algovar{$v$} $=$ \algovar{$\epsilon$}}{\label{alg:approach:returnstatement:begin}
	\algovar{API.return-type} $\leftarrow$ \algovar{$\tau$}\;
	replace \algovar{$s_n$} in \algovar{API.method-body} with \algoval{return $\epsilon$;}\;
}
\Case(\hfill{\commentfont// Variable assignment}){\algovar{$s_n$} : \algovar{$v$} $=$ \algovar{$\epsilon$}}{
	\algovar{API.return-type} $\leftarrow$ \algovar{$\mathcal{T}[v]$}\;
	replace \algovar{$s_n$} in \algovar{API.method-body} with \algoval{return $\epsilon$;}\;
}

\tcc{\hfill\textbf{\pquattro{}}}
\Case{\algovar{$s_n$} : \algoval{System.out.println(string-literal + $\epsilon$)} $\lor$ \algoval{System.out.println($\epsilon$)}}{
	\algovar{API.return-type} $\leftarrow$ \GetTypeOfExp{\algovar{$\epsilon$}, \algovar{imports}, \algovar{classpath}}\;
	replace \algovar{$s_n$} in \algovar{API.method-body} with \algoval{return $\epsilon$;}\;
}
\Other{
	\algovar{API.return-type} $\leftarrow$ \algoval{void}\;
}

\Return{$\langle$\algovar{API}, \algovar{imports}, \algovar{classpath}$\rangle$}\label{alg:approach:returnstatement:end}

\end{algorithm}

\paragraph{Input and output}
\tool takes as an input:
\begin{inparaenum}[(i)]
	\item \algovarref{CS}, a \ac{so} code snippet;
	\item \algovarref{SO-page}, the \ac{so} page of the snippet, which \tool uses to generate the method name;
	\item \algovarref{JARs}, a set of common \java libraries to recover the missing import and variable declarations~\cite{terragni_csnippex_2016}.
\end{inparaenum}
\tool outputs
\begin{inparaenum}[(i)]
	\item \algovarref{API}, the method declaration of \algovarref{CS};
	\item \algovarref{imports}, the import declarations of the non-primitive types that \algovarref{API} references;
	\item \algovarref{classpath}, the libraries in \algovarref{JARs} that declare the types in \algovarref{imports}.
\end{inparaenum}

\paragraph{Preliminary check (\crefrange{alg:approach:getordefault}{alg:approach:wellformed:return})}
\Cref{alg:approach} starts by checking if \algovarref{CS} already contains import declarations (\cref{alg:approach:getordefault}).
If~yes, it extracts them and searches in \algovarref{JARs} for the corresponding libraries, which it adds to the \algovarref{classpath}.
If not (the common case), it creates an empty \algovarref{imports} list and an initial \algovarref{classpath} with only the \ac{jdk} \textsc{JAR} library.
Next, it checks if \algovarref{CS} already defines a well-formed and compilable \api.
If so, it returns \algovarref{CS}, \algovarref{imports}, and \algovarref{classpath} (\crefrange{alg:approach:wellformed:condition}{alg:approach:wellformed:return}), otherwise it starts the \enquote{\apization{}} process.

\paragraph{Initialization of the API (\cref{alg:approach:init})}
The \enquote{\apization{}} process begins by initializing \algovarref{API}, the method declaration for \algovarref{CS}.
By default, the \algovarref{modifiers} of \algovarref{API} are \code{public} (because \apis must be accessible by any other class) and \code{static} (to avoid instantiating objects for invoking the \api). %
The \algovarref{throws-clause} of \algovarref{API} is the generic \code{java.lang.Exception}.
\tool initializes the \algovarref{method-body} of \algovarref{API} with  \algovarref{CS}, the \emph{return type} with \code{void} and the \emph{parameter list} with the empty list.

\paragraph{Method name generation (\cref{alg:approach:namegen})}
For completeness, \tool generates a method name for the \api from the title of the \ac{so} page associated with the code snippet~\cite{yin_learning_2018}. 
Indeed, the title of the \ac{so} page often summarizes the intent of the programming task.
\tool relies on a \acf{pos} Tagger~\cite{spacy} to assign parts of speech (\eg, nouns, verbs, and adjectives) to each word in the title.
Then, \tool creates the method name by combining the main \enquote{verb} of the sentence and the corresponding \enquote{direct object} (\ie, noun).
We consider these two parts of speech because method names are typically verbs or verb phrases.
We do not claim this to be a contribution to this work.
In the future, we plan to investigate state-of-the-art approaches for generating method names~\cite{gao_neural_2019}.

\smallskip
For a statically-typed programming language such as \java, type inference is precise and unambiguous only with compilable code~\cite{mishne_typestatebased_2012}. 
\tool requires complete type information to know the type of the method parameters and return statements.
However, assuming only compilable code is infeasible because most \ac{so} code snippets do not compile~\cite{terragni_csnippex_2016,subramanian_making_2013,phan_statistical_2018}.
\Cref{alg:approach:compile} of \cref{alg:approach} tries to compile the \algovarref{API} (wrapping it in a synthetic \java class) with the current \algovarref{imports} and \algovarref{classpath}.
If any compilation errors arise, \tool attempts to fix them.
Note that, \tool needs to re-compile \algovarref{API} iteratively because fixing a compilation error may reveal others%
~\cite{terragni_csnippex_2016}.
\tool supports two types of compilation errors:
\begin{inparaenum}[(i)]
	\item missing type declarations (\cref{alg:approach:compile:misstypedecl}) and
	\item missing variable declarations (\cref{alg:approach:compile:missvardecl}).
\end{inparaenum}
For other error types \tool terminates (\cref{alg:approach:compile:else}).

\paragraph{Recover missing type declarations (\crefrange{alg:approach:compile:misstypedecl}{alg:approach:compile:misstypedecl:end})}
\tool relies on \csnippex~\cite{terragni_csnippex_2016} to fix missing type declarations. 
\csnippex recovers the import declarations that fix such errors by querying the fully-qualified names of the classes declared in \algovarref{JARs}.
This is challenging because there are often many fully qualified names with the same simple name.
\csnippex addresses the challenge with a greedy algorithm based on the \emph{clustering hypothesis}: \emph{\enquote{the referred library classes in a \java source file often come from the same libraries, and hence their import declarations tend to form clusters that share common package names}}~\cite{terragni_csnippex_2016}.
For example, the code snippet in \cref{fig:study:examples_sogh} (top) leads to two missing type declarations: \code{MessageDigest} and \code{NoSuchAlgorithmException}.
\csnippex identifies the correct import declarations because they share the same package name \code{java.security}.
\csnippex adds the corresponding \java libraries in the classpath and leverages the feedback of the compiler to check if the errors are fixed. %

\paragraph{Recover missing variable declarations (\puno{}, \crefrange{alg:approach:compile:missvardecl}{alg:approach:compile:missvardecl:addparam})}
\tool recovers missing variable declarations to fix the compilation errors and to find matches of \puno{}, which considers undeclared variables as method parameters.
To recover missing variable declarations, 
\tool relies on the \algofuncref{RecoverVarType} function (\cref{alg:approach:compile:missvardecl:recovervartype}).
Given an \api with an undeclared variable~\algovarref{$v$}, this function identifies the most plausible type of \algovarref{$v$} by leveraging the usages of \algovarref{$v$} in the \api, which follows the \textsc{Baker} approach~\cite{subramanian_live_2014}. %

For example, the \ac{so} code snippet in \cref{fig:study:examples_sogh} (top) lacks the declaration of variable \code{tag\_xml}.
\tool correctly infers that the type of \code{tag\_xml} is \code{java.lang.String} because
\begin{inparaenum}[(i)]
	\item the code snippet invokes the method \code{public byte[] getBytes()} using \code{tag\_xml} as the object receiver, and
	\item \code{java.lang.String} declares a method with the same name and return type.
\end{inparaenum}
When there are multiple plausible types, \tool uses a successful compilation as a proxy for correctness.
In fact, \algovarref{API} compiles without errors if the declaration of \code{tag\_xml} has type \code{java.lang.String}.
\Cref{alg:approach:compile:missvardecl:recovervartype} of \cref{alg:approach} also updates \algovarref{imports} and \algovarref{classpath} accordingly, which remain unchanged in our example (the package \code{java.lang} is imported by default).
Next, \tool updates the map \algovarref{\(\mathcal{T}\)}, which stores for each declared variable in \algovarref{CS} its type.
\Cref{alg:approach:compile:missvardecl:addparam} of \cref{alg:approach} adds \code{tag\_xml} as a parameter.
This is the correct parameter, as it was also used by the \ac{gh} developer that performed the manual \apization (\code{tag\_xml} is renamed to \code{s}).

\vspace{2mm}
\medskip
\paragraph{Recognize hard-coded initializations (\pdue{}, \crefrange{alg:approach:hardcodeinit:getloopchangingvars}{alg:approach:hardcodeinit:removebody})}
Function \algofuncref{GetLoopChangingVars} returns the variables \algovarref{LP-VARS} in the method body that have at least one assignment inside a loop (\cref{alg:approach:hardcodeinit:getloopchangingvars} of \cref{alg:approach}).
\pdue{} needs to identify such variables because they will not be considered as parameters.
\Cref{alg:approach:hardcodeinit:forbody} of \cref{alg:approach} scans the statements in \algovarref{API.method-body}  to search for variable initializations that meet the conditions of \pdue{}.
The scan considers the following three statements types:

\textit{1) Variable declaration and initialization \algovarref{$\tau$}~\algovarref{$v$} $=$ \algovarref{$\epsilon$}.}
For example, \code{String findString = "hello"} in \cref{fig:study:examples_sogh} (\algovarref{$\tau$} $=$ \code{String}, \algovarref{$v$} $=$ \code{findString}, and \algovarref{$\epsilon$} $=$ \code{"hello"}).
When \tool encounters such statements, it maps \algovarref{$\tau$} to \algovarref{$v$}, and it adds \algovarref{$v$} to \algovarref{ALREADY-INIT-VARS}, which is a set that maintains the variables that are already initialized.
The function \algofuncref{IsHardCoded} takes in input the type \algovarref{$\tau$} and the expression \algovarref{$\epsilon$} and it returns \algovalref{true} if \algovarref{$\epsilon$} is a hard-coded value, \algovalref{false} otherwise.

If \algovarref{$\tau$} is primitive or \code{String}, the function returns \algovalref{true} if \algovarref{$\epsilon$} does not contain identifiers (\ie, variable, class, method names), \algovalref{false} otherwise. 
Identifiers characterize data dependencies.
For example, \algofuncref{IsHardCoded}(\code{String}, \code{"hello"}) returns \algovalref{true} because \code{"hello"} does not contain identifiers. 

As another example, consider the following code snippet.
\begin{javalst}
String a = "world";
String b = "hello" + a;
\end{javalst}

\algofuncref{IsHardCoded} (\code{String}, \code{"hello" + a}) 
returns \algovalref{false} because \algovarref{$\epsilon$} = \code{"hello" + a} is data dependent to the variable \code{a}.

If \algovarref{$\tau$} is non-primitive, \algovarref{$\epsilon$} must always contain at least one identifier (\code{null} is also an identifier). 
For example the \algovarref{$\epsilon$} of the statement \code{Calendar calendar = Calendar.getInstance();} in \cref{fig:example_first_day} has \code{Calendar} and \code{getInstance} as identifiers.
As such, for non-primitive types, \algofuncref{IsHardCoded} returns \algovalref{true} if \algovarref{$\tau$} is a subclass of \code{java.util.Collection} and after the statement \algovarref{$s_i$} follow $n > 1$ statements that add elements to the collection (\eg, invoke \code{add} methods for \code{java.util.List}, and \code{put} methods for \code{java.util.Map}).
\tool makes a similar consideration for matrices and arrays.

\Cref{alg:approach:hardcodeinit:ishardcoded} \cref{alg:approach} checks if the variable \algovarref{$v$} meets both \pdue{} criteria (\algovarref{$v$} is initialized with a hard coded value and is not a loop variable).
If yes, it adds \algovarref{$v$} of type \algovarref{$\tau$} to the parameter list and removes the declaration statement \algovarref{$s_i$} from the method body.
For example, the statement \code{String findStr = "hello"} at Line~2 in \cref{fig:study:examples_sogh} (bottom) meets both requirements, and thus \tool makes \code{findStr} a method parameter and removes the statement. %

\textit{2) Variable declaration \algovarref{$\tau$}~\algovarref{$v$}.}
These statements are only declarations without initializations.
For such statements, \tool saves the type \algovarref{$\tau$} of \algovarref{$v$} and statement \algovarref{$s_i$}.
\tool needs this information if later it encounters the initialization of \algovarref{$v$}.

\textit{3) Variable assignment \algovarref{$v$} $=$ \algovarref{$\epsilon$}.}
At \cref{alg:approach:hardcodeinit:alreadyinitvars}, \Cref{alg:approach} checks if \algovarref{$v$} belongs to \algovarref{ALREADY-INIT-VARS}.
If yes, \tool skips the statement because it already encountered the initialization of \algovarref{$v$}.
If not, \tool has found the initialization of \algovarref{$v$}.
Then, it updates \textit{ALREADY-INIT-VARS} and checks if the \pdue{} criteria are met.
If yes, it recovers the type of \algovarref{$v$} from \algovarref{\(\mathcal{T}\)} and adds the \algovarref{$v$} to the parameter list.
Then, it removes from the method body both the statement that declares \algovarref{$v$} (\algovarref{\(\mathcal{S}[v]\)}) and the statement that initializes \algovarref{$v$} (\algovarref{$s_i$}).

\paragraph{Check the last statement (\ptre{}, and \pquattro{},  \crefrange{alg:approach:returnstatement:begin}{alg:approach:returnstatement:end})}
At \crefrange{alg:approach:returnstatement:begin}{alg:approach:returnstatement:end}, \Cref{alg:approach} analyzes the last statement (\algovarref{$s_n$}) to decide whether it should be considered as the return statement.

If \algovarref{$s_n$} is a variable declaration or an assignment, then \algovarref{$s_n$} matches \ptre{}, and thus \tool replaces \algovarref{$s_n$} with a statement that returns the expression \algovarref{$\epsilon$}.
\tool recovers the type of \algovarref{$\epsilon$} directly from \algovarref{$s_n$} (if \algovarref{$s_n$} is a declaration) or from \algovarref{\(\mathcal{T}\)} (if \algovarref{$s_n$} is an assignment).

If \algovarref{$s_n$} is an invocation to \code{System.out.println}, then \algovarref{$s_n$}  matches \pquattro{}.
\Cref{alg:approach} extracts the argument \algovarref{$\epsilon$} of the invocation by removing the first string-literal (if it exists), which is likely to represent a placeholder (\eg, \code{System.out.println("result :" + s)}).
Given \algovarref{$\epsilon$}, \cref{alg:approach} recovers \algovarref{$\tau$}, the type of \algovarref{$\epsilon$}, which will be the return type of \algovarref{API}.
Although \code{System.out.println} handles String objects, \algovarref{$\tau$} is not necessarily \code{String}.
In fact, \code{System.out.println(object)} invokes that object's \code{toString()} method to convert the object to a \code{String} representation.
For example, given the last statement \code{System.out.println(count)} in \cref{fig:study:examples_sogh} (bottom), the return type should be \code{int} and not \code{String}.
The function \algofuncref{GetTypeOfExp} analyses \algovarref{$\epsilon$} and \algovarref{classpath} to recover \algovarref{$\tau$}.
If \algovarref{$\epsilon$} is a variable \algovarref{$v$}, the function recovers \algovarref{$\tau$} from the map \algovarref{\(\mathcal{T}[v]\)}.
If \algovarref{$\epsilon$} is a method invocation \algovarref{$m$}, the function consults the declaration of \algovarref{$m$} in \algovarref{classpath} to get its return type.

\section{Evaluation}
\label{sec:evaluation}

This section discusses a series of experiments that we conducted to evaluate \tool. 
In the context of our study, we formulated the following three research questions:
\begin{reqs}
	\item [\req{1}] Does \tool generate \apis that are \textbf{identical} to the ones that a human would produce?
	\item [\req{2}] How effective the \tool algorithm is in identifying the \textbf{method parameters}?
	\item [\req{3}] How effective the \tool algorithm is in identifying the \textbf{return statements}?
	
\end{reqs}

To answer these research questions, we collected a ground truth of human-produced \apis.
We decided not to rely on the \acf{gh} dataset used in \cref{sec:study} to avoid overfitting (\tool is based on the insights extracted from the \ac{gh} dataset).
Instead, we asked \emph{\num{20} human participants to build a ground-truth of \num{200} \apis} by manually performing the \apization of \num{200} \ac{so} code snippets.
All the evaluation data is available in our replication package~\cite{replicationpackage} and published at \url{https://apization.netlify.app/evaluation/}.

\subsection{Evaluation Setup}

\subsubsection{Creating a Collection of \apis from \acl{so}}

We considered the \ac{so} data dump of May 2019~\cite{stackoverflow_dump}, which contains \num{1014980} \ac{so} pages with the tag \java.
From these \ac{so} pages, we selected all the \num{1730251} \ac{so} answer posts with at least one code snippet.

\paragraph{Identifying the compilable \ac{so} code snippets}
We first ran \csnippex on each of the \num{1730251} \ac{so} answer posts, to identify those code snippets for which \csnippex is able to recover the missing type declarations.
\csnippex requires a set of common \java libraries JARs as an input~\cite{terragni_csnippex_2016}.
We obtained such a set by downloading the latest JAR of the top three libraries of each category in the \textsc{Maven Repository}~\cite{maven_repository}.
We then used the dependency resolver of \textsc{Maven} to identify the additional JARs that belong to the runtime dependencies of the selected libraries.
In total, we obtained \num{748} JAR files.
Running \csnippex with a time-budget of \num{5} seconds for each post, it returned compilable \java files for \num{141064} \ac{so} posts.

\begin{table*}[!tb]
	\caption{\req{2} analysis and comparison of the human- (\paramvh{}) and \tool-produced (\paramva{}) parameter lists}
	\label{tab:evaluation:params}
	\centering
	\resizebox{\linewidth}{!}{%
		
\rowcolors{2}{}{gray!10}
\begin{tabular}{
    S[table-format=< 1]
    S[table-format=2]
    S[table-format=2] S[table-format=2.2]
    S[table-format=1.2] S[table-format=1.2] S[table-format=1.2] S[table-format=1.2]
    S[table-format=1.2] S[table-format=1.2] S[table-format=1.2] S[table-format=1.2]
    S[table-format=1.2] S[table-format=1.2] S[table-format=1.2] S[table-format=1.2]
    S[table-format=1.2] S[table-format=1.2] S[table-format=1.2] S[table-format=1.2]
}

\hiderowcolors
\toprule

{\multirow{2}[2]{*}{\textbf{\makecell{Param.\\$\abs{\text{\paramvh}}$}}}} & {\multirow{2}[2]{*}{\textbf{\makecell{Human\\\apis}}}} & \multicolumn{2}{c}{\textbf{$\text{\paramvh} \equiv \text{\paramva}$}} & \multicolumn{4}{c}{\textbf{$\abs{\text{\paramvh} \setminus \text{\paramva}}$}} & \multicolumn{4}{c}{\textbf{$\abs{\text{\paramvh} \cap \text{\paramva}}$}} & \multicolumn{4}{c}{\textbf{$\abs{\text{\paramva} \setminus \text{\paramvh}}$}} & \multicolumn{4}{c}{\textbf{\acf{jd}}} \\
\cmidrule(lr){3-4} \cmidrule(lr){5-8} \cmidrule(lr){9-12} \cmidrule(lr){13-16} \cmidrule(lr){17-20}
& & {\textbf{Count}} & {\textbf{\si{\percent}}} & {\textbf{Mean}} & {\textbf{Min}} & {\textbf{Mdn}} & {\textbf{Max}} & {\textbf{Mean}} & {\textbf{Min}} & {\textbf{Mdn}} & {\textbf{Max}} & {\textbf{Mean}} & {\textbf{Min}} & {\textbf{Mdn}} & {\textbf{Max}} & {\textbf{Mean}} & {\textbf{Min}} & {\textbf{Mdn}} & {\textbf{Max}} \\

\midrule
\showrowcolors

0 & 58 & 45 & 77.59 & {--} & {--} & {--} & {--} & {--} & {--} & {--} & {--} & 0.36 & 0.00 & 0.00 & 5.00 & 0.22 & 0.00 & 0.00 & 1.00 \\
1 & 93 & 60 & 64.52 & 0.32 & 0.00 & 0.00 & 1.00 & 0.68 & 0.00 & 1.00 & 1.00 & 0.13 & 0.00 & 0.00 & 2.00 & 0.34 & 0.00 & 0.00 & 1.00 \\
2 & 35 & 7 & 20.00 & 1.14 & 0.00 & 1.00 & 2.00 & 0.86 & 0.00 & 1.00 & 2.00 & 0.29 & 0.00 & 0.00 & 2.00 & 0.58 & 0.00 & 0.50 & 1.00 \\
\geq 3 & 14 & 1 & 7.14 & 2.86 & 0.00 & 3.00 & 6.00 & 0.64 & 0.00 & 0.00 & 4.00 & 0.21 & 0.00 & 0.00 & 1.00 & 0.82 & 0.00 & 1.00 & 1.00 \\

\hiderowcolors
\midrule

{Total ($\geq 0$)} & 200 & 113 & 56.50 & 0.77 & 0.00 & 0.50 & 6.00 & 0.72 & 0.00 & 1.00 & 4.00 & 0.23 & 0.00 & 0.00 & 5.00 & 0.38 & 0.00 & 0.00 & 1.00 \\

\bottomrule

\end{tabular}

	}
\end{table*}

\paragraph{Creating the \ac{so} \apis}
We  ran \tool on these \num{141064} \ac{so} answer posts with a time budget of \num{10} seconds each, obtaining \textbf{\num{109930} \apis}.
\tool skipped \num{31134} out of the \num{141064} posts because the \apization is either impossible or ambiguous.
It is impossible for abstract methods and for \java files with only field or class declarations. 
It is ambiguous for files that have more than one public method or that declare more than one class.
In such cases, \tool cannot infer which public method is the intended \api.

\begin{revised}
It is worth noting that, for each of the produced \apis, \tool generates a \textsc{JavaDoc} containing the link to the original \ac{so} post from which the code was taken (see~\cref{fig:example_first_day}).
This is compliant with the \ac{so} Terms of Service, which, at present, states that user contributions are licensed under \emph{Creative Commons Attribution-ShareAlike}\footnote{\url{https://stackoverflow.com/legal/terms-of-service/public\#licensing}}.
The specific license terms depend on the date of publication of the \ac{so} post, but all of them require appropriate credit to the authors of the content, \ie, a link to the \ac{so} post.
In fact, the \emph{CC BY-SA} license allows re-distribution and re-use of a licensed work (even for commercial use) on the condition that the creator is appropriately credited.
However, it is the responsibility of the end user to keep the link of the \ac{so} post associated with the \tool-generated \apis. 
Similarly, manually copying and adapting a \ac{so} snippet should require appropriate credit by including a link to the \ac{so} post~\cite{baltes_attribution_2017}.
\end{revised}

\subsubsection{Selecting the \apis for the Evaluation}
\label{subsec:evaluation:selecting_apis}

From the \num{109930} \apis we selected those that satisfy five properties: %

\smallskip
\noindent
\textbf{I.} The \ac{so} page of the \api is a \query{how to} question.
Following previous \ac{so} studies, we assume that the most useful code snippets are in answers to \query{how to} questions~\cite{treude_how_2011,treude_understanding_2017}.
We identified such questions by the presence of the word \query{how} in the \ac{so} page title~\cite{treude_how_2011}.

\smallskip
\noindent
\textbf{II.} The \ac{so} post associated with the \api is the accepted answer or has a score of at least two (two is the average score in \ac{so}).
This is to select high-quality code snippets.

\smallskip
\noindent
\textbf{III.} The \ac{so} post associated with the \api contains exactly one code snippet.
This is to avoid ambiguity, as multiple code snippets in the same \ac{so} post often refer to alternative solutions of the same programming task.
Having only one code snippet, the human participant does not need to decide which one to consider.

\smallskip
\noindent
\textbf{IV.} The import declarations of the \api do not refer to any external libraries other than the \ac{jdk}.
Participants might produce incorrect \apizations, for instance, if they are unfamiliar with a particular library.

\smallskip
\noindent
\textbf{V.} The \ac{so} code snippet associated with the \api \emph{does not} contain a well-formed method declaration.
In such cases, the code snippet is already an \api,
and \cref{alg:approach} has no effect.

\smallskip
A total of \num{9901} \apis satisfy all of these properties.
We sorted them by the view count of the corresponding \ac{so} post and selected the first \num{200} \apis.
It is worth noting that we had to manually discard some of the \apis in which the \apization is not a reasonable operation (even though the above-mentioned properties were satisfied).
For example, when the \ac{so} code snippet is not a programming task (\eg, it shows usage examples of JDK classes), or it is semantically incomplete (\eg, it contains placeholders for missing functionality).
The \num{200} \apis have \num{11.45} lines of code on average.
The corresponding \ac{so} posts have an average number of views of $\approx$\num{66000}, and an average score of \num{46.62}.

\subsubsection{Ground-Truth of Human \apizations}

We partitioned the \num{200} code snippets in \num{200} disjoint sets and sent them to \num{20} expert \java developers in the authors' circle of acquaintances.
Each participant had assigned ten \ac{so} posts.
The \num{20} participants come from seven different countries and constitute a heterogeneous group of ten Ph.D. students majoring in software engineering, five senior software engineering researchers, and five professional \java developers.
The participants have several years of experience in \java programming: \num{9.8} years on average (min \num{1}, median \num{9.5}, and max \num{15}).
None of the \num{20} participants knew that \tool exists and how it generates \apis.
Thus, they performed the manual \apization without biases.

\paragraph{Experiment description}
Each participant received a script that interacts via the command line.
The script gives the instructions and monitors the \apization time. 
It was an uncontrolled experiment, thus they ran the script at their convenient time.
We decided to avoid guidelines to let the participants decide what \apization means to them.
Instead, the script exemplifies the concept with an example.
After showing the example, the script shows the \ac{so} page of the first assigned code snippet.
It then asks the participant to read the \ac{so} page to understand the semantics of the code snippet, and to write in the IDE a method declaration for it.
This process repeats until the participant completes the ten assigned code snippets.
This led to \num{200} pairs $\langle\text{\apivh}, \text{\apiva}\rangle$ of human- (\apivh) and \tool-produced (\apiva) \apis from the same \ac{so} code snippet.
We release the instructions of the script in our replication package~\cite{replicationpackage} and published at \url{https://apization.netlify.app/evaluation/script/}.

\paragraph{Pre-processing the human \apis}
Before comparing the pairs, we inspected the \num{200} human-produced \apis to fix any compilation errors and to check whether the participants renamed any parameters.
We corrected one compilation error, and we renamed the parameters of \num{27} human \apis to match the ones automatically generated by \tool.
We also removed, from \num{15} human-produced \apis, variable declarations for return statements that \tool avoids by construction.
For example, \code{int a = b + c; return a;} becomes \code{return b + c;}.

\subsection{\req{1}: Identical \apis}

To check for identical \apis, we compared each pair $\langle\text{\apivh}, \text{\apiva}\rangle$ with the state-of-the-art source code differencing tool \textsc{GUMTREE}~\cite{falleri_finegrained_2014}.
When comparing the pairs, we excluded differences in method names.
\textsc{GUMTREE} implements an \ac{ast} differencing algorithm that takes into account fine-grained \ac{ast} differences while ignoring irrelevant differences in the source code, \ie, new lines, white spaces, and comments.

\begin{figure}[!tb]
	\centering
	\includegraphics[width=0.95\linewidth, trim=0 5 0 0, clip]{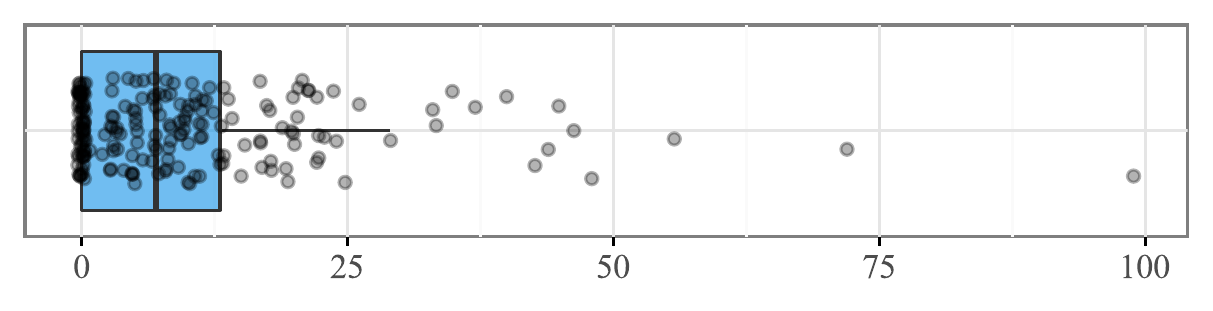}
	\caption{Distribution of the number of \ac{ast} differences.}
	\label{fig:study:ast_differences_boxplot}
\end{figure}

\Cref{fig:study:ast_differences_boxplot} shows the distribution of the number of \ac{ast} differences of the \num{200} pairs, which ranges from \numrange{0}{99} (average \num{9.85} and median \num{7}).
Interestingly, \num{63} (\SI{31.50}{\percent}) \apis generated by \tool are identical to the human-produced ones ($\langle\text{\apivh}, \text{\apiva}\rangle$ has zero \ac{ast} differences).
The pair in \cref{fig:example_first_day} is one of such identical \apizations in our experiments.

\begin{revised}
Achieving identical \apizations is an unrealistic expectation, as in some cases, the participants modified the method body of the \api by removing \code{System.out.println} statements or unnecessary variables.
\req{2} and \req{3} give more insights about the dissimilar pairs by studying the \tool effectiveness in extracting the parameters and return statements while ignoring superficial differences in the method bodies.
\end{revised}

\begin{custombox}{\req{1} -- In summary}
	\begin{revised}
	\tool generated \num{63} (\SI{31.50}{\percent}) \apis identical (including the method-body and import declarations) to the human-produced ones.
	\end{revised}
\end{custombox}

\subsection{\req{2}: Method Parameters}

To answer \req{2}, we extracted and compared the parameter lists of the \num{200} pairs.
Given a pair $\langle\text{\apivh}, \text{\apiva}\rangle$, we use \paramvh and \paramva to denote the parameter lists of \apivh and \apiva, respectively. 
Note that the order of elements in the parameter list is irrelevant, thus we considered \paramvh and \paramva as unordered sets.
For example, for the \api pair of \cref{fig:example_first_day}, $\text{\paramvh} = \text{\paramva} = \{\text{\code{int week}}, \text{\code{int year}}\}$.

\Cref{tab:evaluation:params} breaks down the human-produced \apis (\apivh) by the number of parameters (the cardinality of \paramvh).
The participants produced \num{58} \apis without parameters, and \num{142} \apis with one or more parameters (Column \enquote{Human \apis} of \cref{tab:evaluation:params}).
The rest of \cref{tab:evaluation:params} compares \paramvh with the corresponding \paramva. 

Column \enquote{$\text{\paramvh} \equiv \text{\paramva}$} of \cref{tab:evaluation:params} indicates the number and percentage of \apis pairs with equivalent \paramvh and \paramva.
\paramvh and \paramva are equivalent if they are both empty, or contain identical parameters.
Two parameters $p_h \in \text{\paramvh}$ and $p_a \in \text{\paramva}$ are identical if and only if they
\begin{inparaenum}[(i)]
	\item have the same type;
	\item have the same identifier, \ie, variable name;
	\item refer to the same variable in the method body.
\end{inparaenum}
For example, in the pair of~\cref{fig:example_first_day}, the parameters \code{int week} \paramvh and \code{int week} in \paramva are identical.
They have the same type and identifier, and the two bodies refer to them in the same way.
\tool generates \num{113} (\SI{56.50}{\percent}) \apis with equivalent parameter lists to the human-produced ones ($\text{\paramvh} \equiv \text{\paramva}$).
\begin{revised}
When the human-produced \apis have two or more parameters, the number of equivalent pairs decreases. 
This is an expected result.
Intuitively, the more parameters the manually-crafted ground truth \api has, the harder it is for \tool to extract an identical parameters list.
It is worth mentioning that, in principle, there is no difference if an \api has one or more parameters.
This is because \cref{alg:approach} considers each variable in the code snippet individually.
In practice, we observed that the majority of human-produced \apizations have at most one parameter.
We observed this situation both in the \num{135} \apis used for extracting the patterns and the \num{200} \apis used to evaluate \tool (\cref{tab:evaluation:params}).
In fact, the average number of parameters of the \num{135} \apis is \num{0.33}. 
The reason for that could be that code snippets often target atomic operations that require one input only.
\end{revised}

Column \enquote{$\abs{\text{\paramvh} \setminus \text{\paramva}}$} of \cref{tab:evaluation:params} shows descriptive statistics (mean, min, median, and max) of the number of missing parameters for each \api pair (when $\abs{\text{\paramvh}} \geq 1$).
Intuitively, $\abs{\text{\paramvh} \setminus \text{\paramva}}$ indicates the number of parameters in \paramvh missing from the corresponding \paramva.
The value ranges from \numrange{0}{6} with an average of \num{0.77} and a median of \num{0.50}.
Among the \num{142} \apis with $\abs{\text{\paramvh}} \ge 1$, \num{68} of them (\SI{47.88}{\percent}) have zero missing parameters ($\abs{\text{\paramvh} \setminus \text{\paramva}} = 0$).

Column \enquote{$\abs{\text{\paramvh} \cap \text{\paramva}}$} of \cref{tab:evaluation:params} indicates the number of parameters in common between each \api pair (when $\abs{\text{\paramvh}} \geq 1$).
The value ranges from \numrange{0}{4} with an average of \num{0.72} and a median of \num{1.00}.
Among the \num{142} \apis with $\abs{\text{\paramvh}} \ge 1$, \num{91} of them (\SI{64.08}{\percent}) have at least one parameter in common ($\abs{\text{\paramvh} \cap \text{\paramva}} \ge 1$).
This indicates that \tool often identifies the same parameters that a human would identify.

Column \enquote{$\abs{\text{\paramva} \setminus \text{\paramvh}}$} of \cref{tab:evaluation:params} shows the number of spurious parameters for each \api pair (those extracted by \tool, but not by the human participants).
The value ranges from \numrange{0}{5} with an average of \num{0.23} and a median of \num{0.00}.
Among the \num{200} \apis, \num{166} of them (\SI{83.00}{\percent}) do not have spurious parameters ($\abs{\text{\paramva} \setminus \text{\paramvh}} = 0$).
This demonstrates that \tool seldom extracts parameters that a human would not extract.

Column \enquote{\acf{jd}} of \cref{tab:evaluation:params} reports the \emph{Jaccard Distance}~\cite{small_cocitation_1973} between \paramvh and \paramva, and it is defined as $\jd\left(\text{\paramvh}, \text{\paramva}\right) = \frac{\abs{\text{\paramvh} \cap \text{\paramva}}}{\abs{\text{\paramvh} \cup \text{\paramva}}}$ from \numrange{0}{1}.
The lower the value is, the more similar the two sets are.
If \paramvh and \paramva are both empty, $\jd\left(\text{\paramvh}, \text{\paramva}\right)$ returns \num{0.0}.
The values range from \numrange{0.00}{1.00} with an average of \num{0.38} and a median of \num{0.00}.
These results confirm that in most cases, humans and \tool extracted identical parameter lists.
Notably, for nine parameters \tool and the humans extracted the same variables but inferred compatible albeit different types. 
For example, \code{java.util.Collection} and \code{java.util.List}, \code{double} and \code{int}. %
In such cases we consider the parameters to be different.

\begin{custombox}{\req{2} -- In summary}
	\begin{revised}
	\tool generated \num{113} (\SI{56.50}{\percent}) \apis with identical parameter lists to the human-produced ones.
	\end{revised}
 \end{custombox}

\subsection{\req{3}: Return Statements}

\begin{table}[!tb]
	\caption{\req{3} return statements comparison}
	\label{tab:evaluation:return}
	\centering
	\resizebox{\linewidth}{!}{%
		\setlength{\tabcolsep}{3mm}
\rowcolors{2}{}{gray!10}
\begin{tabular}{
    l l S[table-format=2] S[table-format=2.2]
    S[table-format=2] S[table-format=3.2]
}
\hiderowcolors
\toprule

\multicolumn{4}{c}{\textbf{Return Type}} & \multicolumn{2}{c}{\textbf{\makecell{Equivalent Return Type\\and Statements}}} \\
\cmidrule(lr){1-4} \cmidrule(lr){5-6}
\textbf{\apivh} & \textbf{\apiva} & {\textbf{Count}} & {\textbf{\si{\percent}}} & {\textbf{Count}} & {\textbf{\si{\percent}}} \\

\midrule
\showrowcolors

\texttt{void} & \texttt{void} & 63 & 31.50 & 63 & 100.00 \\
\texttt{void} & not \texttt{void} & 2 & 1.00 & {--} & {--} \\
not \texttt{void} & \texttt{void} & 72 & 36.00 & {--} & {--} \\
not \texttt{void} & not \texttt{void} & 63 & 31.50 & 52 & 82.54 \\

\hiderowcolors
\midrule

{Total} & & 200 & & 115 & \\

\bottomrule

\end{tabular}

	}
\end{table}

\Cref{tab:evaluation:return} breaks down the \num{200} \apis pairs by return types (\code{void} and not \code{void}).
Column \enquote{Equivalent Return Type and Statements} %
counts the number and percentage of \apis with equivalent return statements.
A pair of \apis $\langle\text{\apivh}, \text{\apiva}\rangle$ has equivalent return statements if
\begin{inparaenum}[(i)]
	\item both \apis have \code{void} as return type; or
	\item both \apis return the same type and have identical return statements in the method body.
\end{inparaenum}
\num{115} (\SI{57.50}{\percent}) of the \num{200} \apis pairs have equivalent return statements.
This indicates that \tool can effectively identify the return type and statements that a human would identify. 

When both the human and \tool added a return statement (row not \code{void}, not \code{void} in \cref{tab:evaluation:return}), \SI{82.54}{\percent} of times they used the same type and return statements.
This indicates that the conservative nature of our algorithm leads to few spurious return statements.

 \begin{custombox}{\req{3} -- In summary}
	\begin{revised}
	\tool generated \num{115} (\SI{57.50}{\percent}) \apis with identical return statements to the human-produced ones.
	\end{revised}
 \end{custombox}

\subsection{Discussion}

Our experimental results are both promising and encouraging. %
Indeed, for \num{163} (\SI{81.50}{\percent}) \apis generated by \tool, either the return statements or method parameters were the same as those produced by the developers.
Note that a \ac{so} code snippet could have more than one plausible \api.
Some of the \apis obtained by \tool could be plausible albeit different from the manually-produced ones.
Thus, our experimental setup only under-approximates the effectiveness of \tool.

\paragraph{Comparing \apization efforts}
The average \apization time for the participants ranges from \SI{17}{\second} to \SI{15}{\minute} and \SI{58}{\second}, with an average of \SI{4}{\minute} and \SI{22}{\second}, and a median of \SI{3}{\minute} and \SI{22}{\second}.
Note that the participants performed the task offline without our supervision.
As such, we cannot tell if a participant was distracted during the experiment.
However, these values give an idea of the order of magnitude of the manual effort required.
Regarding the \num{200} code snippets of this experiment, the average execution time of \cref{alg:approach} was $\approx$\SI{8}{\second} for each code snippet.
This shows the potential usefulness of \tool in reducing software development costs.
Considering that developers re-use code from \ac{so} several times in one day~\cite{storey_evolution_2014}, \tool could help speed up the software development process.

\paragraph{\begin{revised}{False negatives}\end{revised} due to literals as parameters}
We investigated why some pairs of \apis were different, identifying one main reason (\num{39} cases): \emph{literals-as-parameters}, \emph{when strings and number literals in the arguments of method calls become parameters}.

For example, consider the \apization in \url{https://apization.netlify.app/evaluation/comparison/8192887/}.
Both the human and \tool extracted \code{list} as parameter, but the human also extracted the \code{String} literal \code{bea} from \code{string.matches("(?i)(bea).*")}.

\tool adopts a conservative approach that tolerates missing parameters but minimizes spurious ones, as the results of \req{2} demonstrate.
We could have designed \tool to extract all strings and number literals in the method body.
Although this would yield fewer false negatives, it would also lead to more spurious parameters since not all string and number literals should become parameters.

\begin{revised}
We believe that it is better to have false negatives rather than false positives when extracting parameters.
This is because extracting literals from the method body \enquote{removes} information, which has to be recovered from the \ac{so} code snippet.
For example, consider the code snippet in \cref{fig:study:examples_sogh} (top).
\tool does not extract the string-literal \code{MD5} as a parameter. 
Indeed, any random string yields incorrect code.
If \code{MD5} was extracted, the user would need to recover the missing value \code{MD5} from the \ac{so} code snippet.
Correctly recognizing and handling the literal-as-parameter issue is an important future work as it will drastically reduce the false negatives of \tool.
\end{revised}

\begin{revised}
\paragraph{Maintainability of the \apis}
Currently, \tool returns a dedicated class for each generated \apis.
The end users are free to import the class as it is or copy and paste the method and import declarations inside their codebases.
Indeed, having many one-method classes results in less cohesive software and ultimately negatively impacts the system's quality.
An essential future work would be to propose a technique to group semantically related \tool-generated \apis into the same \java class.
For instance, one could group \apis that import the same classes and take as input the same type of parameters (\eg, strings, lists, arrays).
This will lead to a library of \tool-generated \apis more similar to a manually-written \api, facilitating the search, use, and maintainability of \apis automatically extracted from \ac{so}.
\end{revised}

\subsection{Threats to Validity}

\paragraph{Threats to internal validity}

A possible threat to internal validity is the choice of the \num{200} code snippets for the evaluation.
They might not be a representative sample of code snippets. We tried to mitigate such a risk by selecting a reasonably large number of snippets for an evaluation involving human participants.
Furthermore, by selecting popular code snippets, \ie, based on the views count, we ensured that we selected a relevant sample.

\paragraph{Threats to external validity}
A possible threat to external validity is that the four patterns %
are specific to \java, and might not generalize well for other programming languages.
For instance, in the case of dynamically-typed languages like \textsc{Python}, the \apization is easier for some aspects but harder for others.
On the one hand, it is difficult to identify possible parameters and return statements by relying on the types of literals.
On the other hand, the flexibility of dynamic types allows extracting parameters easier than a statically-typed language like \java.
Repeating our study for dynamically-typed languages is an important future work.

\begin{revised}
Another threat to the external validity is that currently \tool only handles two types of compilation errors: missing type declarations and missing variable declarations.
\tool cannot produce \apis for those code snippets that have other types of compilation errors. 
However, these two types are among the most common compilation errors in \ac{so} code snippets~\cite{terragni_csnippex_2016}. 
\tool relies on previous techniques (\csnippex~\cite{terragni_csnippex_2016} and \textsc{Baker}~\cite{subramanian_live_2014}) to fix compilation errors. 
In the future, \tool could rely on other techniques to handle additional types of compilation error.
For instance, a common compilation error in \ac{so} code snippets is \code{compiler.err.expected}~\cite{terragni_csnippex_2016}, which means the code does not comply with the syntax rules of the \java language. 
Examples of such rules are: \enquote{a semicolon should be at the end of every statement, or there should be a matching sequence of opening and closing brackets.}
\tool could rely on a parser that recognizes and fixes such errors.
\end{revised}

\paragraph{Threats to construct validity}
A possible threat to construct validity relates to the metrics that we used to evaluate \tool.
We measured the effectiveness of \tool by counting how many times \tool and the humans made the same \apization choices.
However, a \ac{so} code snippet could have more than one plausible \ac{api}.
Additional  human evaluators could help recognize when \tool generated a plausible \ac{api}, albeit different from the human-produced one.
Nevertheless, we preferred to rely on a objective method, even if it might have resulted in a disadvantage for \tool, but is not biased by a subjective evaluation.

\section{Related Work}
\label{sec:related_work}

\acf{so} provides an important source of crowd-generated data that inspired and powered many techniques and tools. %
In a recent systematic mapping study, Meldrum \etal identified \num{266} research papers that rely on \ac{so} data to accomplish various software engineering taks~\cite{meldrum_crowdsourced_2017}.
It includes topics like program repair~\cite{liu_mining_2018}, mobile development issues~\cite{vasquez_exploratory_2013,rosen_what_2016,beyer_manual_2014,kavaler_using_2013}, \apis misuses and issues~\cite{zhang_are_2018,wang_recommending_2015,ahasanuzzaman_classifying_2018,vasquez_how_2014}, and technology landscape discovery~\cite{chen_techland_2016,chen_mining_2016}. 
In this paper, we propose \tool to facilitate the reuse and analysis of \ac{so} code snippets by transforming them into compilable and reusable \apis.
To the best of our knowledge, it is the first attempt to accomplish this.
In the following, we discuss the most related work in code snippet analysis, search, and reuse.

\paragraph{Code snippet analysis}
Recently, Terragni \etal proposed \csnippex to resolve compilation errors of \ac{so} code snippets~\cite{terragni_csnippex_2016}.
\tool leverages this tool to resolve type declaration errors.
Subramanian and Holmes studied the compilability of \ac{so} code snippets~\cite{subramanian_making_2013}.
However, in the case of missing method declarations, these approaches simply wrap the code snippets in a synthetic method.
Differently from \tool, they do not aim at %
identifying the method parameters and return statements of code snippets. 

Researchers have proposed to mine intent-snippet pairs for code summarization or search~\cite{yao_staqc_2018,iyer_summarizing_2016,wong_autocomment_2013,yang_query_2016,zagalsky_example_2012,yin_learning_2018}.
The intent of the snippet is often characterized by the \ac{so} question title~\cite{yin_learning_2018}.
These techniques analyze the code snippets to identify which lines of code are related to the \ac{so} title while filtering out all the implementation details.
\tool has the opposite goal of generating the missing implementation details to make the code snippet easy to invoke.
All of these techniques aim at identifying the lines of code associated with the intent and do not aim to generate a proper method declaration for the extracted lines of code.
\tool could work in synergy with these techniques by creating an \api for the code extracted by these techniques.

\paragraph{Code snippet search}
There is also a large body of work on improving code search in on-line resources (such as \ac{so})~\cite{mishne_typestatebased_2012, mcmillan_exemplar_2012, mcmillan_portfolio_2013}.
A popular approach to facilitate search of \ac{so} code is to 
reduce the context switching from \textsc{IDEs} (\eg, \textsc{IntelliJ IDEA} and \textsc{Eclipse}) to web browsers by incorporating \ac{so} code search into \textsc{IDEs}. %
\textsc{Prompter}~\cite{ponzanelli_mining_2014} and \textsc{Seahawk}~\cite{ponzanelli_seahawk_2013} recommend \ac{so} posts into the IDE based on source code context found in the \textsc{IDE}.
\textsc{T2API}~\cite{nguyen_t2api_2016}, \textsc{NLP2Code}~\cite{campbell_nlp2code_2017}, and \textsc{RACK}~\cite{rahman_rack_2017} recommend code snippets extracted from \ac{so} based on natural language text describing the programming task.
\textsc{RACK} leverages crowd-source knowledge taken from both \ac{so} and \textsc{GitHub}.
\textsc{StackInTheFlow}~\cite{greco_stackintheflow_2018} improves the previous approaches by monitoring the behavior of the developers to personalize the retrieved posts.
All of these techniques aim to improve the code search or reduce the context switching from \textsc{IDEs} to browsers.
Differently from \tool, they do not help developers to integrate the \ac{so} code snippet into their code base.
\tool complements such approaches, as it could extract, compile and create \apis for the code snippets that are retrieved by these techniques.

\paragraph{Code snippet reuse}
Zhang \etal~\cite{zhang_are_2018}  proposed \textsc{ExampleStack}, a \textsc{Google Chrome} extension that highlights in a \ac{so} page the statements that were changed when a \ac{gh} developer previously reused the same code snippet.
Such highlights help developers to adapt the code snippet in their code bases.
To know which statements should be highlighted, \textsc{ExampleStack} queries an archive of \ac{so} code reuses in \ac{gh} projects.
Zhang \etal built such an archive by analyzing \num{200} code reuses across \ac{so} code snippets and \ac{gh} projects.
Similarly to \tool, \textsc{ExampleStack} aims at facilitating the adaptation of code snippets, but with completely different goals.
First, \textsc{ExampleStack} suggests general code changes~\cite{zhang_are_2018}.
Differently from \tool, it does not automatically extract method parameters and return statements, and it does not aim to generate compilable \apis.
Second, \textsc{ExampleStack} can suggest changes for only those code snippets present in the precomputed archive. Conversely, \tool does not require any prior knowledge on the code snippet under analysis.
Third, the input of \textsc{ExampleStack} and \tool differs substantially.
\textsc{ExampleStack} analyzes parsable code snippets with a well-defined method declaration~\cite{zhang_are_2018}, or by wrapping the snippets with synthetic method headers%
~\cite{subramanian_making_2013}.
Instead, \tool analyzes incomplete code snippets.

\section{Conclusion and Future Work}
\label{sec:conclusions}

Online developers forums like \acf{so} have drastically changed how developers write code~\cite{vasilescu_how_2014,abdalkareem_what_2017,ye_structure_2017,brandt_examplecentric_2010,storey_evolution_2014,mao_survey_2017}.
Developers constantly visit \ac{so} for finding solutions to programming tasks.
The \ac{so} revolution has been recognized by the software engineering community and several techniques have been proposed to facilitate the reuse and analysis of \ac{so} code snippets~\cite{mishne_typestatebased_2012, mcmillan_exemplar_2012, mcmillan_portfolio_2013}. 

In this paper, we presented \tool, an approach that transforms \ac{so} code snippets into compilable and reusable \apis.
To the best of our knowledge, this is new to \ac{so} code snippet analysis.
Our empirical results demonstrate the usefulness of \tool in reducing the developers' effort and enabling the creation of a large dataset of \apis from \ac{so}.

There are several possible future works, and we highlight the three most promising ones.
First, address the literal-as-parameter issue by employing machine learning to recognize which literal should become a parameter.
Second, %
investigate state-of-the-art approaches~\cite{nguyen_suggesting_2020,liu_learning_2019,gao_generating_2020,gao_neural_2019}  to generate semantically meaningful method names.
In particular, in our case, one could generate method names by relying on both the natural language free text in the \ac{so} posts (\eg, the discussions and comments) and the code snippet itself.
Third, explore text summarization and code comment generation approaches~\cite{kim_enriching_2013,wong_autocomment_2013,treude_augmenting_2016,parnin_measuring_2011} to generate the \textsc{JavaDoc}. %

\bibliography{references, urls}

\end{document}